\documentclass[
aps,pre,onecolumn, a4paper, notitlepage,superscriptaddress]{revtex4-2}

\usepackage{xcolor}
\usepackage[hidelinks]{hyperref}
\hypersetup{colorlinks=false, linktoc=all}

\usepackage{graphicx}
\usepackage{comment}
\usepackage[caption=false]{subfig}

\usepackage{amsmath}
\usepackage{amssymb}
\usepackage{mathtools}

\usepackage{amsthm}
\newtheorem{theorem}{Theorem}
\newtheorem{claim}{Claim}

\theoremstyle{definition}
\newtheorem*{Clustering}{Clustering}
\newtheorem*{DecisionClustering}{Decision Clustering}

\usepackage{physics}
\usepackage[ruled,vlined,linesnumbered]{algorithm2e}

\DeclareMathOperator{\N}{\mathbb{N}}
\DeclareMathOperator{\R}{\mathbb{R}}

\DeclareMathOperator*{\argmin}{\arg\min}

\DeclareMathOperator*{\zo}{\{0,1\}}

\newcommand{\dist}[2]{\mathrm{dist}(#1, #2)}
\newcommand{\deff}{d_{\mathrm{eff}}}

\begin{document}

\title{Quantum density peak clustering}

\author{Duarte Magano}
	\affiliation{Instituto Superior T\'{e}cnico, Universidade de Lisboa, Portugal}
	\affiliation{Instituto de Telecomunica\c{c}\~{o}es, Portugal}
\author{Lorenzo Buffoni}
	\affiliation{Portuguese Quantum Institute, Portugal}
\author{Yasser Omar}
	\affiliation{Instituto Superior T\'{e}cnico, Universidade de Lisboa, Portugal}
	\affiliation{Portuguese Quantum Institute, Portugal}
    \affiliation{Centro de Física e Engenharia de Materiais Avançados (CeFEMA), Physics of Information and Quantum Technologies Group, Portugal}

\date{July 21, 2022}

\begin{abstract}
Clustering algorithms are of fundamental importance when dealing with large unstructured datasets and discovering new patterns and correlations therein, with applications ranging from scientific research to medical imaging and marketing analysis.
In this work, we introduce a quantum version of the density peak clustering algorithm, built upon a quantum routine for minimum finding.
We prove a quantum speedup for a decision version of density peak clustering depending on the structure of the dataset.
Specifically, the speedup is dependent on the heights of the trees of the induced graph of nearest-highers, i.e., the graph of connections to the nearest elements with higher density.
We discuss this condition, showing that our algorithm is particularly suitable for high-dimensional datasets.
Finally, we benchmark our proposal with a toy problem on a real quantum device.
\end{abstract}

\maketitle

\section{Introduction}

Machine Learning (ML) \cite{bishop_pattern_2011,hastie2009elements} is a field with an exceptional cross-disciplinary breadth of applications and studies. The aim of ML is to develop computer algorithms that improve automatically through experience by learning from data, so as to identify distinctive patterns and make decisions with minimal human intervention. The applications of ML that are already possible today are extremely compelling and diverse \cite{sutton2018reinforcement,graves2013speech,sebe2005machine}, and still growing at a steady pace. However, the training and deployment of these models, involving an ever increasing amount of data, faces computational challenges \cite{tieleman2008training} that are only partially met by the development of special purpose classical computing units such as GPUs.

This challenge posed by ML algorithms has led to a recent interest in applying quantum computing to machine learning tasks \cite{schuld2015introduction,wittek2014quantum,adcock2015advances,arunachalam2017survey,Biamonte:2017db} that sparked the field of Quantum Machine Learning (QML). So far, there have been many proposals for different QML algorithms. Several of them \cite{lloyd2020quantum,vinci2020path,otterbach2017unsupervised} have shown the potential to accelerate ML tasks, but largely rely on heuristic methods, and \emph{proving} an advantage in terms of computational complexity for these particular algorithms is generally hard. On the other hand, there have been a number of works that, by using quantum algorithms with known complexity as subroutines \cite{wiebe2012quantum, lloyd2014quantum}, could prove the existence of a quantum advantage in QML. Indeed, it has been conjectured that QML algorithms could provide the first breakthrough algorithms on near-term quantum devices, given the inherent robustness of these algorithms to noise and perturbations.

Most of the literature on QML has focused on \emph{supervised} learning \cite{goodfellow2016deep} problems. This particular subset of ML has a number of appealing characteristics as it is more immediate to implement and allows for feedback loops and minimization of well-kown and well-behaved cost functions. Unlike supervised learning, unsupervised learning is a much harder, and still largely unsolved, problem. And yet, it has the appealing potential to learn the hidden statistical correlations of large unlabeled datasets \cite{vincent2008extracting,hinton1995wake}, which constitute the vast majority of data being available today. 

Amongst all unsupervised learning problems, clustering is one of the most popular.
In clustering, we consider a dataset $D$ composed of $n$ elements,
\begin{equation}
D = \{ x_0, \ldots, x_{n-1}\},
\end{equation}
in which we are given a notion of \emph{distance} between every two elements in the dataset $x_i, x_j \in D$,
\begin{equation}
\dist{x_i}{x_j}.
\end{equation}
Interpreting this distance as a similarity measure, the (ill-defined) problem of clustering can be formulated as follows.
\begin{Clustering}
Given a dataset $D$ and a distance $\dist{\cdot}{\cdot}$ between each pair of elements of $D$, partition $D$ into sets called \emph{clusters}, such that similar elements belong to the same cluster and dissimilar elements belong to distinct clusters.
\end{Clustering}

Clustering algorithms thus need to separate unlabeled data into different classes (or clusters) without any external labeling and supervision. 
Solutions to the clustering problem based on quantum computing have been proposed for a long time, resorting to a plethora of different strategies \cite{Aimeur2007,Yu2010,Li2011,Aimeur2013,LloydMohseniRebentrost,otterbach2017unsupervised,Bauckhage2017,Daskin2017,qmeans,li2021quantum,QSpectralClustering,pires2}.
For example, in Ref. \cite{Li2011} the data poins are treated as interacting quantum walkers on a lattice; whereas in Ref. \cite{qmeans} quantum subroutines for distance estimation and matrix arithmetics are employed to develop an efficient quantum version of a classical standard algorithm, $k$-means clustering.
In \cite{otterbach2017unsupervised}, the clustering problem was reformulated as an optimisation problem and solved by applying a hybrid optimisation algorithm on a Rigetti quantum processor, hinting to the possibility of realising such advantage on near term quantum devices.

In this work, we adopt the black-box/oracular model of clustering introduced in \cite{Aimeur2007}, in which the information concerning the distances between points in the dataset is available only through oracle queries.
Under this framework, references \cite{Aimeur2007,Aimeur2013}, using variants of Grover's search \cite{GroverSearch}, quantize typical subroutines of learning algorithms, such as finding the largest distance in a dataset, computing the median, or constructing the $c$-neighbourhood graph.
These subroutines are then used to accelerate standard clustering methods, namely, divisive clustering, $k$-medians clustering and clustering via minimum spanning tree.
Nevertheless, the state-of-the-art in clustering has been steadily evolving in recent years, and so the question arises: can we also use quantum computing to speedup modern clustering algorithms?

We consider a recent and very popular clustering method, usually referred to as density peak clustering (DPC) \cite{Rodriguez2014}.
In a nutshell, the idea is to attribute a ``density'' value to every element of the dataset based on their distances to all other elements, and then assign each element to the same cluster as the nearest neighbour that has a density greater than itself (referred to as its \emph{nearest-higher}).
Relying on a variant of quantum search known as quantum minimum finding \cite{DurrHoyer1996}, we show that, for any element of the dataset, we can find its nearest-higher (up to bounded-error probability)in time $\mathcal{O}\left(n^{3/2}\right)$, as opposed to the classical $\mathcal{O}\left(n^2\right)$ complexity.
Unfortunately, to fully solve DPC we would need to repeat this subroutine $n$ times.
This would provide no advantage as we can classically implement DPC in $\mathcal{O}\left(n^2\right)$ time.

Motivated by this, in this work we consider a simpler variant of the clustering problem, which we refer to a \emph{decision} clustering.
\begin{DecisionClustering}
Given a dataset $D$, a distance $\dist{\cdot}{\cdot}$ between each pair of elements of $D$, and two elements $x_i, x_j \in D$, decide if $x_i$ and $x_j$ are in the same cluster.
\label{prob:decisionclustering}
\end{DecisionClustering}
\noindent
Evidently, any solution of the clustering problem contains the answer to decision clustering, but not the other way around. 
This problem is relevant in situations where one is interested in establishing connections between particular elements but does not need to know the cluster structure of the entire dataset. 
For example, we may want to know if two users of social media platform are friends, or if an individual fits a particular consumer segment.
Moreover, this can also be useful when one has to decide to which cluster a new datapoint belongs without having to re-cluster all the data.
Finally, decision clustering can be iteratively applied to cluster small subsets of the data.

Classically solving decision clustering with DPC has the same complexity as solving the full clustering problem, $\mathcal{O}\left(n^2\right)$.
However, in the quantum setting we show that we can solve it in $\tilde{\mathcal{O}} \left(n^{3/2} H \right)$ time, where $H$ is the maximum height of all trees in the graph of nearest-highers (to be properly defined in Section \ref{sec:DPC}).
The factor $H$ is not known \emph{a priori}, depending on the specific structure of the dataset.
Nevertheless, we argue that for high-dimensional datasets $H$ scales as $\mathcal{O}(n^{1 / \deff})$, where $\deff$ is a constant greater than $2$, and confirm this hypothesis with numerical simulations.
When this holds, quantum density peak decision clustering provides a speedup over its classical counterpart.

Finally, we benchmark our quantum algorithm using a toy problem on a real quantum device, the ibm-perth 7-qubit quantum processor.
As expected, the hardware errors severally mitigate any possible quantum advantage.  
Nevertheless, the noiseless simulations confirm the potential of our approach.

The article is structured as follows.
Section \ref{sec:preliminaries} provides summary of background material that is necessary for understanding our quantum algorithm. 
Namely, we introduce the data model and review the quantum minimum finding algorithm.
Section \ref{sec:DPC} explains the classical DPC algorithm, assuming that the reader is not yet familiar with it.
Then, in section \ref{sec:qalg} we present our quantum algorithm.
The complexity of the algorithm depends on the $H$ factor, and so in section \ref{sec:height} we study how $H$ behaves for different datasets.
In section \ref{sec:experiment} we show the results of the implementation of our proposal on a real quantum device.
Section \ref{sec:conclusion} concludes the article with a brief summary and discussion.

\section{Preliminaries} 
\label{sec:preliminaries}

\subsection{Data model}

In this article, we work in a black-box model.
We assume that our knowledge about the data comes uniquely from querying an ``oracle'' that returns the distance between pairs of points
\begin{equation}
	(i,j) \xrightarrow[]{\text{query}} \dist{x_i}{x_j}.
\label{eq:querymodel}
\end{equation}
We make no assumptions about this distance besides that it is non-negative and symmetric. 
That is, it does not need to be a distance by a proper mathematical definition.
Our complexity measure is the number of performed queries (this is known as query complexity).

This model is an abstraction that reasonably fits a number of problems.
For example, the oracle may represent accessing a database with the distances between a group of cities, or a routine that estimates the dissimilarity between pairs of images.
The query complexity is a good estimate of the total time complexity whenever determining the distance between elements is the most computationally intensive part of the algorithm.
Nevertheless, we also point out that this model may not be a good description of other common situations.
For example, if we know the coordinates of a set of points in $\R^d$ (for some integer $d$), we already have more structure than in the black-box model and there are geometrical methods that allow significant speedups for certain tasks.

In the quantum setting, we assume that we can query the distances between elements in quantum superposition.
That is, we assume access to a unitary $Q$ (the quantum oracle) such that
\begin{equation}
	Q \ket{i,j,0^{\otimes q}} = \ket{i, j, \dist{x_i}{x_j}},
\label{eq:quantumquerymodel}
\end{equation}
where $q$ is the number of bits necessary to store the distances up to desired accuracy.
In particular, given a superposition $\sum_{i j} \alpha_{ij} \ket{i, j}$ (for any set of normalized complex amplitudes $\{\alpha_{ij}\}_{ij}$), $Q$ acts as
\begin{equation}
	Q \bigg( \sum_{i j} \alpha_{ij} \ket{i, j} \bigg) \ket{0^{\otimes q}} 
	=
	\sum_{i j} \alpha_{ij} \ket{i, j, \dist{x_i}{x_j}}.
\end{equation}
The quantum query complexity is counted as the number of applications of the unitary $Q$ (for more details on the quantum query model refer to \cite{Ambainis_2017}).
When describing classical data, the oracle may be realized with a quantum random access memory (qRAM) architecture \cite{QRAM}.
In the end, our results are critically dependent on the existence of a qRAM with the above mentioned properties, representing a common setting in theoretical work on quantum algorithms, and in quantum machine learning in particular.
Nevertheless, even though there have been proposals of physical architectures for implementing QRAM \cite{QRAM_architectures,Park_Petruccione_Rhee_2019}, there are still significant challenges to overcome before such a device can be practically realized \cite{Matteo_Gheorghiu_Mosca_2020}.

\subsection{Quantum minimum finding \label{sec:quantumminimumfinding}}

The key quantum routine we use in our work is the well-known quantum minimum finding algorithm of D\"{u}rr and H\o{}yer \cite{DurrHoyer1996}, which in itself is a specific application of the quantum search algorithm \cite{GroverSearch}.

Begin by considering a boolean function $F$ defined on a domain of size $n$, $F: \{0, 1, \ldots, n-1\} \rightarrow \zo$.
Our goal is to find an element $x$ such that $F(x)=1$, assuming one exists.
$F$ is provided as a black box, that is, we can only gain information about the function by evaluating it on given elements.
In the worst-case scenario, we may need to query all $n$ possible inputs before succeeding.

Now suppose that we have access to a unitary $O_F$ that marks the $1$-inputs with a $-1$ phase,
\begin{equation}
O_F \ket{i} = 
\begin{cases}
+\ket{i}, \text{ if } F(i) = 0 \\
-\ket{i}, \text{ if } F(i) = 1
\end{cases}.
\label{eq:unitaryF}
\end{equation}
The unitary $O_F$ is an oracle for $F$, and an application of $O_F$ is referred to as a quantum query to $F$ as introduced in the section above.
Lets now consider $\ket{\Psi}$ to be the uniform superposition of all input states,
\begin{equation}
	\ket{\Psi} = \frac{1}{\sqrt{n}} \sum_{i=0}^{n-1} \ket{i}.
\end{equation}
If we measure $\ket{\Psi}$, we obtain every input with equal probability.
Grover's algorithm \cite{GroverSearch} is based on the observation that we can amplify the probability of measuring $1$-input states via the repeated application of the operator
\begin{equation}
	G_F = \left(2 \ket{\Psi}\bra{\Psi} - I \right) \cdot O_F
\end{equation}
to an initial state $\ket{\Psi}$.
Roughly, the amplitudes of the $1$-input states grow linearly with each application of $G_F$, while their measurement probabilities grow quadratically.
This means that $\sim\sqrt{n}$ quantum queries are sufficient to measure an input state with high probability.
Boyer, Brassard, H\o{}yer, and Tapp \cite{Boyer1996}, generalizing Grover's algorithm \cite{GroverSearch}, prove the following theorem.
\begin{theorem}[Quantum search, \cite{Boyer1996}] \label{thm:quantumsearch}
Let $F:\{0, \ldots, n-1\} \rightarrow \zo$ and let $t = \vert \{x \in \{0, \ldots, n-1\} : F(x)=1\} \vert$ (which does not need to be known a priori).
Then, we can find an element $x$ such that $F(x) = 1$ with an expected number of $\mathcal{O}(\sqrt{n / t})$ quantum queries to $F$.
\end{theorem}

Quantum minimum finding calls quantum search as a subroutine.
Suppose that we want to find a minimizer of a black-box function $f: \{0, \ldots, n-1\} \rightarrow \N$.
For any element $i \in [n]$, define the boolean function
\begin{equation}
F_i(j)=
\begin{cases}
0, \text{ if } f(j) \geq f(i) \\
1, \text{ if } f(j) < f(i)
\end{cases},
\end{equation}
which can be evaluated with two queries to $f$.
Let $O_{F_i}$ be a unitary that evaluates $F_i$, as in expression \eqref{eq:unitaryF}.
The quantum minimum finding algorithm starts by choosing a threshold element $i$ uniformly at random between $0$ and $n-1$.
Employing quantum search with $O_{F_i}$ as oracle, we look an element $j$ such that $F(j) < F(i)$. 
We then repeat this process, updating $j$ as the threshold element, until the probability that the selected threshold is the minimum of $f$ is sufficiently large.
With this algorithm, D\"{u}rr and H\o{}yer \cite{DurrHoyer1996} reach the following result.
\begin{theorem}[Quantum minimum finding, \cite{DurrHoyer1996}] \label{thm:quantumminimumfinding}
Let $f:\{x \in \{0, \ldots, n-1\} \rightarrow \N$ and $\epsilon \in [0, 1[$.
Then, we can find the minimizer of $f$ with probability at least $1 - \epsilon$ using $\mathcal{O}(\sqrt{n} \log(1 / \epsilon) )$ quantum queries to $f$.
\end{theorem}

\section{Density peak clustering \label{sec:DPC}}

Since its introduction in 2014  \cite{Rodriguez2014}, density peak clustering (DPC) has been widely studied and applied \cite{tu2019spatial,cheng2016large,shi2019unsupervised}.
This algorithm, albeit remaining quite simple in the concept and implementation, presents some interesting features that are absent from most of the simpler clustering algorithms discussed in the QML literature. 
As an example, it does not make assumptions on the number of clusters present in the data (unlike simpler algorithms such as $k$-means \cite{macqueen1967some}), but it can infer this information from the data itself. 
Linked to this first property, DPC does not require any prior hypothesis on the shape of the dataset (i.e., gaussian distributed or symmetric), and works well with datasets of virtually any shape \cite{Rodriguez2014,fang2020adaptive}. 
Finally, DPC is able to detect outliers in the dataset, that is, elements that do not belong to any cluster. 
This last property opens the possibility of using DPC beyond the usual scope of clustering problems, such as anomaly detection \cite{tu2020hyperspectral,shi2019unsupervised}. 
We will now introduce this algorithm from a mathematical point of view and have a deeper look at its implementation to understand its capabilities and its criticalities.

In density peak clustering, the first step is to compute the density $\rho(x_i)$ of every element $x_i \in D$,
\begin{equation}
	\rho(x_i) = \sum_{x_j \in D} \chi(\dist{x_i}{x_j}),
\label{eq:densitydefinition}
\end{equation}
where $\chi$ is a convolutional kernel that can be optimized according to the specific application.
Common choices include the step kernel $\chi(x) = \Theta(x - d_c)$ or the Gaussian kernel $\chi(x) = \exp(- x^2 / d_c^2)$, for some normalization parameter $d_c$.
Next, for each element $x_i$ we find its ``nearest-higher'' $h(x_i)$, defined as the closest element to $x_i$ with higher density than $x_i$,
\begin{equation}
	h(x_i) = \argmin_{x_j: \rho(x_j) > \rho(x_i)} \dist{x_i}{x_j}.
\label{eq:NHdefinition}
\end{equation}
The nearest-higher separation $\delta(x_i)$ is the distance between $x_i$ and its nearest-higher,
\begin{equation}
	\delta(x_i) = \dist{x_i}{h(x_i)}.
\label{eq:deltadefinition}
\end{equation}
Naturally, if the point in question is the one with highest density in the entire dataset then it does not have a nearest-higher.
In that case, by convention, the nearest-higher separation is $+\infty$.

\begin{figure}[p]
\centering
\begin{subfloat}[]{\label{fig:clustering_illustration}
	\includegraphics[width=0.42\linewidth]{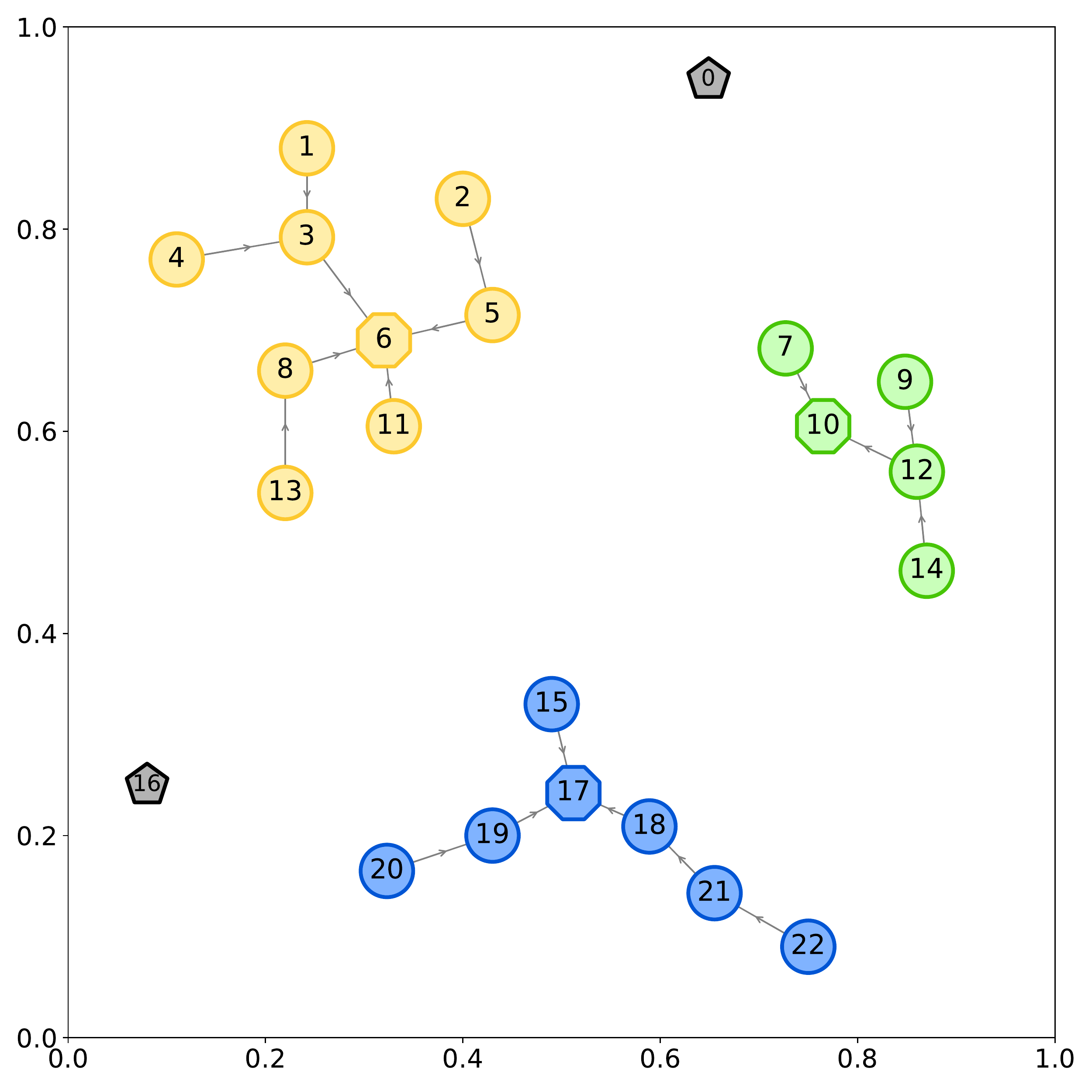}
	}
\end{subfloat}
\begin{subfloat}[]{\label{fig:rhodelta_plot}
	\includegraphics[width=0.42\linewidth]{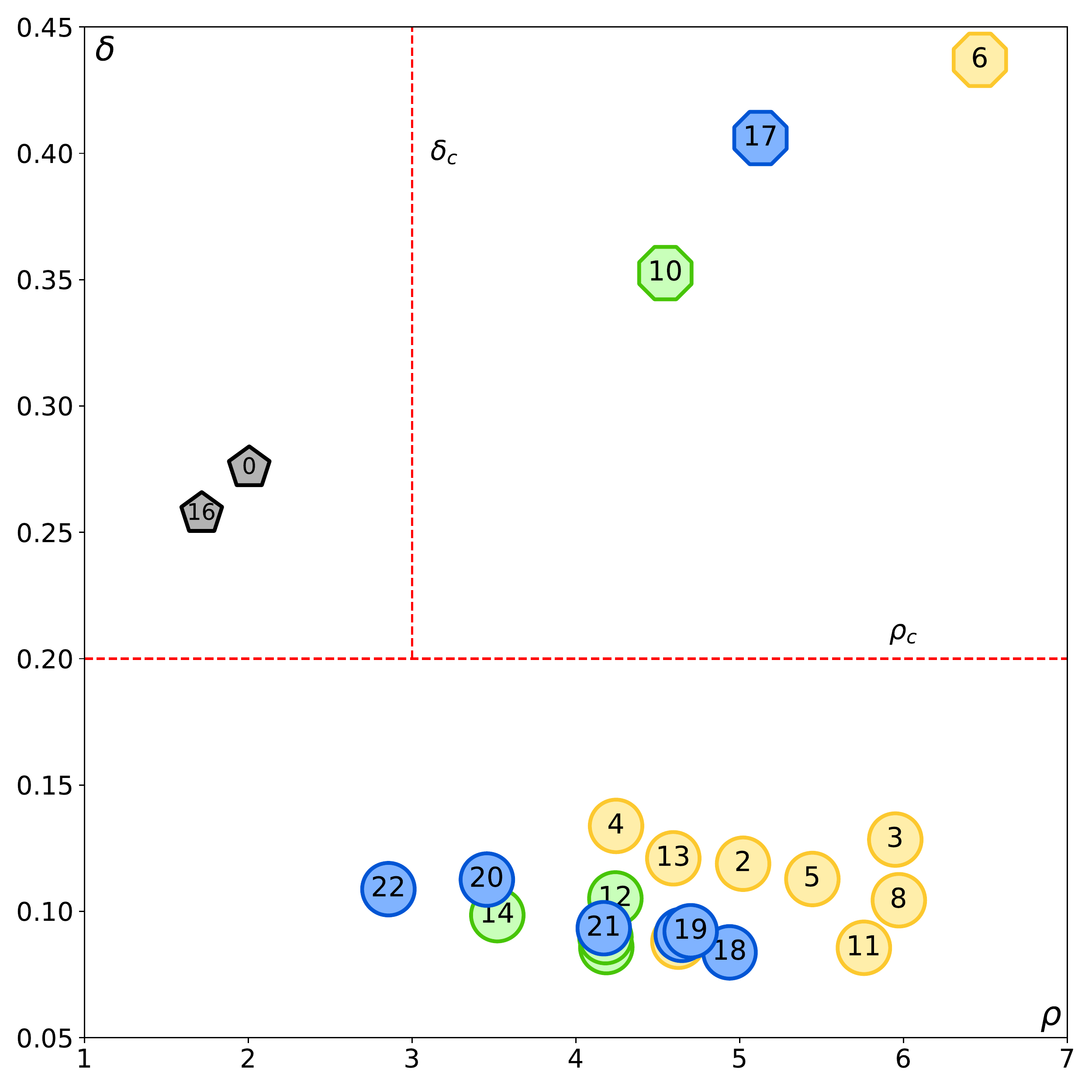}
	}
\end{subfloat}
\caption{Density peak clustering. 
	In Figure \ref{fig:clustering_illustration} we represent a dataset of $23$ elements embedded in $\R^2$, with the similarity measure provided by the Euclidean distance. 
	The roots are represented by octagons, the outliers by pentagons, and the rest of the dataset by circles.
	The elements are coloured by cluster (with the outliers coloured in black).
	The gray arrows indicate the edges of the directed graph of nearest-higher connections.
	This graph forms a forest, with one tree per distinct cluster.
	In Figure \ref{fig:rhodelta_plot} we plot the nearest-higher separation as a function of density for all points in the dataset.
	The elements satisfying $\rho(x_i) > \rho_c \land \delta(x_i) > \delta_c$ are promoted to roots, 
	while the ones satisfying $\rho(x_i) < \rho_c \land \delta(x_i) > \delta_c$ are demoted to outliers.
	}
\label{fig:DPCillustration}
\end{figure}

The key observation at the basis of DPC is that, for the elements that are local maxima of the density,  the nearest-higher separation is much larger than the typical nearest-neighbour distance.
This idea is illustrated in Figure \ref{fig:DPCillustration}, where we consider a small dataset embedded in $\R^2$ with the similarity measure provided by the Euclidean distance.
In Figure \ref{fig:rhodelta_plot}, by plotting the density versus the nearest-higher separation for all elements in the dataset, it becomes clear that the elements $7$, $11$, and $15$ stand out from their neighbours.
These three elements are classified as \emph{roots}, and
each root will originate its own cluster.
The elements $0$ and $16$, despite also having large nearest-higher separations, also have low densities and so are classified as \emph{outliers}.
In more detail, for some choice of threshold $\rho_c$ and $\delta_c$ (to be specified according to the typical scales of the dataset), we promote to roots the elements $x_i$ satisfying
\begin{equation}
	\rho(x_i) > \rho_c \quad \text{and} \quad \delta(x_i) > \delta_c,
\end{equation}
and demote to outliers those who obey
\begin{equation}
	\rho(x_i) < \rho_c \quad \text{and} \quad \delta(x_i) > \delta_c.
\end{equation}
Then, the rest of the elements are assigned to the same cluster as their nearest-higher.
As evidenced in Figure \ref{fig:clustering_forest}, the directed graph of nearest-highers (with edges originated from roots removed) forms a forest and each tree corresponds to a distinct cluster, the root nodes of the tree being the elements that were prometed to roots.

The main steps of density peak clustering are summarized in Algorithm \ref{algo:DPclustering}. 
The complexity of the algorithm becomes clear immediately from the first step.
Indeed, in order to compute the density of a given element, we need to query the distance to every other element in the dataset.
Repeating this for all elements in the dataset means that we end up querying all $(n^2 - n) / 2$ distances.
\begin{claim}
The density peak clustering algorithm (Algorithm \ref{algo:DPclustering}) has $\mathcal{O}(n^2)$ query complexity, where $n$ is the size of the dataset.
\label{thm:DDclassicalcomplexity}
\end{claim}

Now consider a \emph{decision} version of the clustering problem, where the goal is, given two elements $x_i, x_j \in D$, to decide if they belong to the same cluster.
Evidently, we could solve the full clustering problem and then verify if $x_i$ and $x_j$ were assigned to the same cluster.
But in the context of DPC there is a more direct approach, relying on the following simple observation: the two elements belong to the same cluster if and only if the sequences of nearest-highers starting from $x_i$ and $x_j$ lead to the same root.
In other words, the problem is reduced to finding the roots of the respective trees in the graph of nearest-highers.
This can be solved with Algorithm \ref{algo:decisionclustering}, where we just walk up the trees node by node by computing the corresponding nearest-higher.
We know that we have reached a root when the nearest-higher separation is larger than $\delta_c$.

Unfortunately, this approach comes with no significant complexity advantage as every step we take up the tree has essentially the same computational cost of the full clustering.
To see this, just note that computing a nearest-higher of an element $x_i$ requires knowing how the density of all other element compares to $\rho(x_i)$ (\textit{cf.} Eq.~\eqref{eq:NHdefinition}).
So, it seems unavoidable to compute all the densities.
\begin{claim}
The decision version of density peak clustering (Algorithm \ref{algo:decisionclustering}) has $\mathcal{O}(n^2)$ query complexity, where $n$ is the size of the dataset.
\label{thm:DDclassicaldecisioncomplexity}
\end{claim}

In the next section, we show that we can circumvent this cost with quantum search, establishing a quantum advantage for the decision version of density peak clustering.

\begin{algorithm}[p]
\SetArgSty{}
\caption{Density peak clustering (originally proposed in Ref. \cite{Rodriguez2014})}
\label{algo:DPclustering}
\SetKwInOut{Input}{input}\SetKwInOut{Output}{output}
\Input{dataset $D$, density threshold $\rho_c$, and nearest-higher separation threshold $\delta_c$ (and possibly normalization parameter $d_c$, depending on the form of the kernel)}
\Output{clusters}
For all $x_i \in D$, compute density $\rho(x_i) = \sum_{j} \chi(\dist{x_i}{x_j})$ \label{step:densities}\;
For all $x_i \in D$, compute nearest-higher $h(x_i) = \argmin_{j: \rho(x_j) > \rho(x_i)} \dist{x_i}{x_j}$ and respective separation $\delta(x_i) = \dist{x_i}{h(x_i)}$\label{step:NHs}\;
For all $x_i \in D$, promote to root if $\rho(x_i) > \rho_c \land \delta(x_i) > \delta_c$ and to outlier if $\rho(x_i) < \rho_c \land \delta(x_i) > \delta_c$\label{step:promotions}\;
For all $x_i \in D$ that is not a root nor an outlier, assign $x_i$ to the same cluster as $h(x_i)$\label{step:assignclusters}\;
\end{algorithm}

\begin{algorithm}[p]
\SetArgSty{}
\caption{Decision version of density peak clustering}
\label{algo:decisionclustering}
\SetKwInOut{Input}{input}\SetKwInOut{Output}{output}
\SetKwData{no}{no} \SetKwData{yes}{yes}
\SetKwFunction{findroot}{FindRoot}
\SetKwProg{Fn}{def}{\string:}{}
\Input{dataset $D$, $x_i, x_j \in D$, density threshold $\rho_c$, and nearest-higher separation threshold $\delta_c$ (and possibly normalization parameter $d_c$, depending on the form of the kernel)}
\Output{yes/no}
\Fn{\findroot{$x$}}{
	\If{$x$ is a root}{
		output $x$\;
	}
	\Else{
		output \findroot{$h(x)$}\;
	}
}
\If{$x_i$ or $x_j$ are outliers}{
	output \no\;
}
\If{\findroot{$x_i$} = \findroot{$x_j$}}{
	output \yes\;
}
\Else{
	output \no\;
}
\end{algorithm}

\section{Quantum algorithm \label{sec:qalg}}

We now have all the elements to introduce a quantum algorithm to solve the decision version of density peak clustering. The quantum algorithm will use the same strategy as in Algorithm \ref{algo:decisionclustering}, calling the quantum minimum finding routine (section \ref{sec:quantumminimumfinding}) to determine the nearest-highers.

For each $i \in \{0, \ldots, n-1\}$, consider the function
\begin{align}
f_i(j)
&=
\begin{cases}
\dist{x_i}{x_j} , & \text{if } \rho_j > \rho_i\\
+ \infty, & \text{if } \rho_j \leq \rho_i.
\end{cases}
\label{eq:ffunction}
\end{align} 
From the definition of nearest-higher (Eq. \eqref{eq:NHdefinition}), it is clear that finding the minimizer of $f_i$ is equivalent to determining the nearest-higher of $x_i$,
\begin{equation}
h(x_i) = \argmin_j f_i(j).
\label{eq:NH}
\end{equation}
For any element in the dataset, we can evaluate its density (Eq.~\eqref{eq:densitydefinition}) with $n-1$ distance queries.
So, we can compute $f_i(j)$, for any  $j \in \{0, \ldots, n-1\}$, with $\mathcal{O}(n)$ queries.
Since any classical computation can be simulated on a quantum computer \cite{NielsenChuang}, there exists a unitary $U_i$ that uses $\mathcal{O}(n)$ quantum queries and achieves the transformation
\begin{equation}
U_i \ket{j} \ket{0^{\otimes q + 1 }} =  \ket{j} \ket{f_i(j)}
\end{equation}
for any $j \in \{0, \ldots, n-1\}$, where $q$ is the number of qubits to store the distance up to the desired accuracy and we add an extra flag qubit to indicate if $f_i$ evaluates to $+\infty$.
Therefore, by Theorem \ref{thm:quantumminimumfinding}, we can use the quantum minimum finding algorithm to find the minimizer of $f_i$ (i.e., the nearest-higher of $x_i$) with probability at least $1 - \epsilon$ with $\mathcal{O}(\sqrt{n} \log (1 / \epsilon))$ applications of $U_i$, amounting to a total of 
\begin{equation}
	\mathcal{O}\left(n^{3/2} \log (1 / \epsilon)\right)
\label{eq:NHquantumcomplexity}
\end{equation}
quantum queries.

Let $H$ be the maximum height of any tree in the graph of nearest-highers.
In the worst case scenario, one may need to call quantum minimum finding $2 H$ times before reaching the root node for both inputs.
Moreover, we want to make sure that it finds the correct nearest-highers every single call with high probability, which can be done by setting the constant $\epsilon$ in equation \eqref{eq:NHquantumcomplexity} to be sufficiently small.
A simple calculation concludes the following.
\begin{claim}
Let $D$ be a dataset of $n$ elements and let $H$ be the maximum height of any tree in the graph of nearest-highers.
For any two elements $x_i, x_j \in D$, we can solve density peak decision clustering with success probability at least $1 - \epsilon$ (for any $\epsilon > 0$) with quantum query complexity
\begin{equation}
\mathcal{O} \left( n^{3/2} H \log (H / \epsilon) \right).
\end{equation}
\end{claim}

Recall that the classical complexity of the decision version of density peak clustering is $\mathcal{O}(n^2)$ (Claim \ref{thm:DDclassicaldecisioncomplexity}), irrespective of the value of $H$.

\section{Height of trees \label{sec:height}}

We have seen that we can reach a lower quantum query complexity for the density peak decision clustering problem depending on the factor $H$, the maximum height of the trees in the graph of nearest-highers.
Indeed, there is quantum speedup when $H$ scales as $\mathcal{O}(n^a)$ for some $a<1/2$.
In contrast, there is no speedup when $H = \Omega(n^a)$ with $a > 1/2$, as we may solve clustering classically in $\mathcal{O}(n^2)$ time. 

\begin{figure}[t]
\centering
\begin{subfloat}[]{\label{fig:clustering_ball}
	\includegraphics[height = 19em]{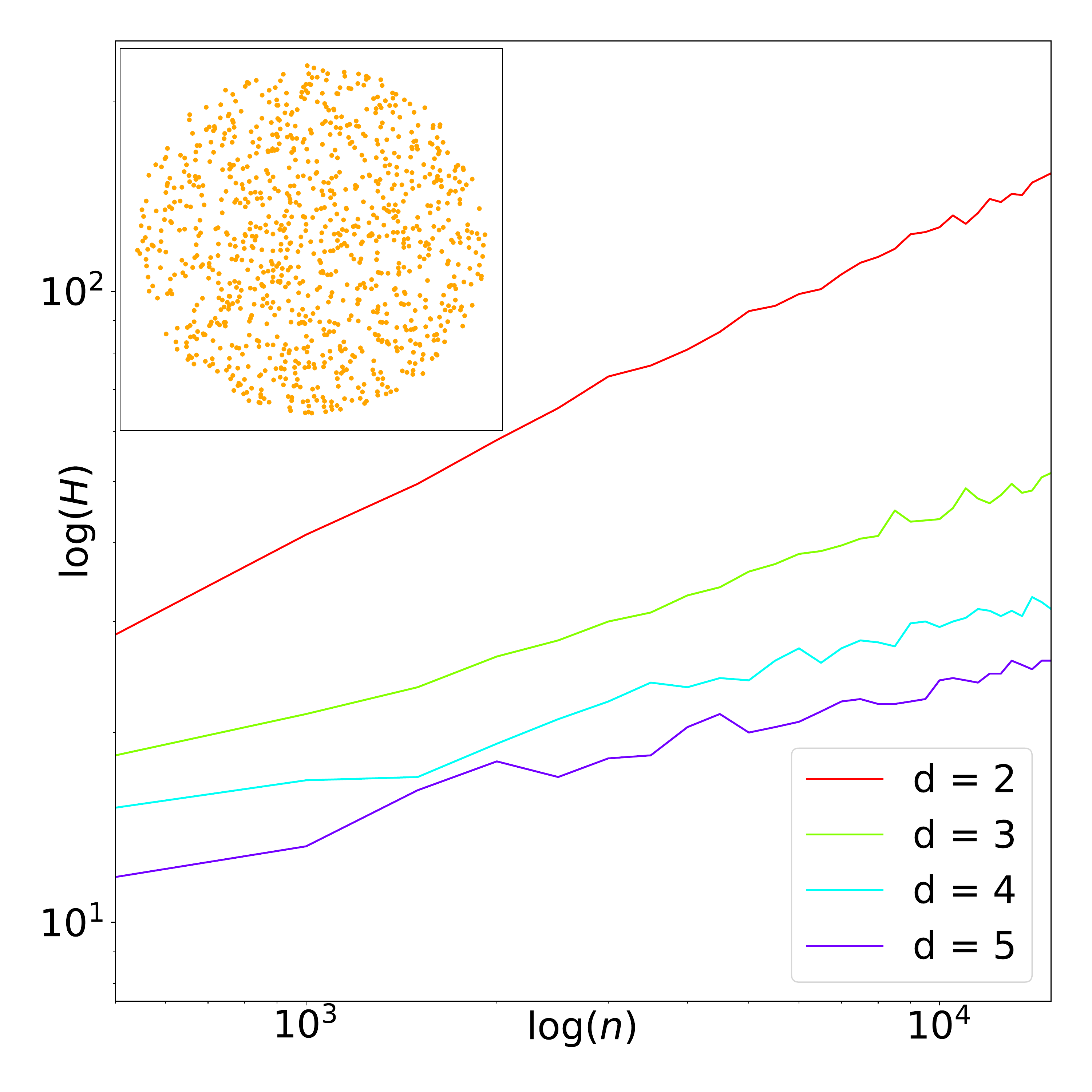}
	}
\end{subfloat}
\begin{subfloat}[]{\label{fig:clustering_gaussian}
	\includegraphics[height = 19em]{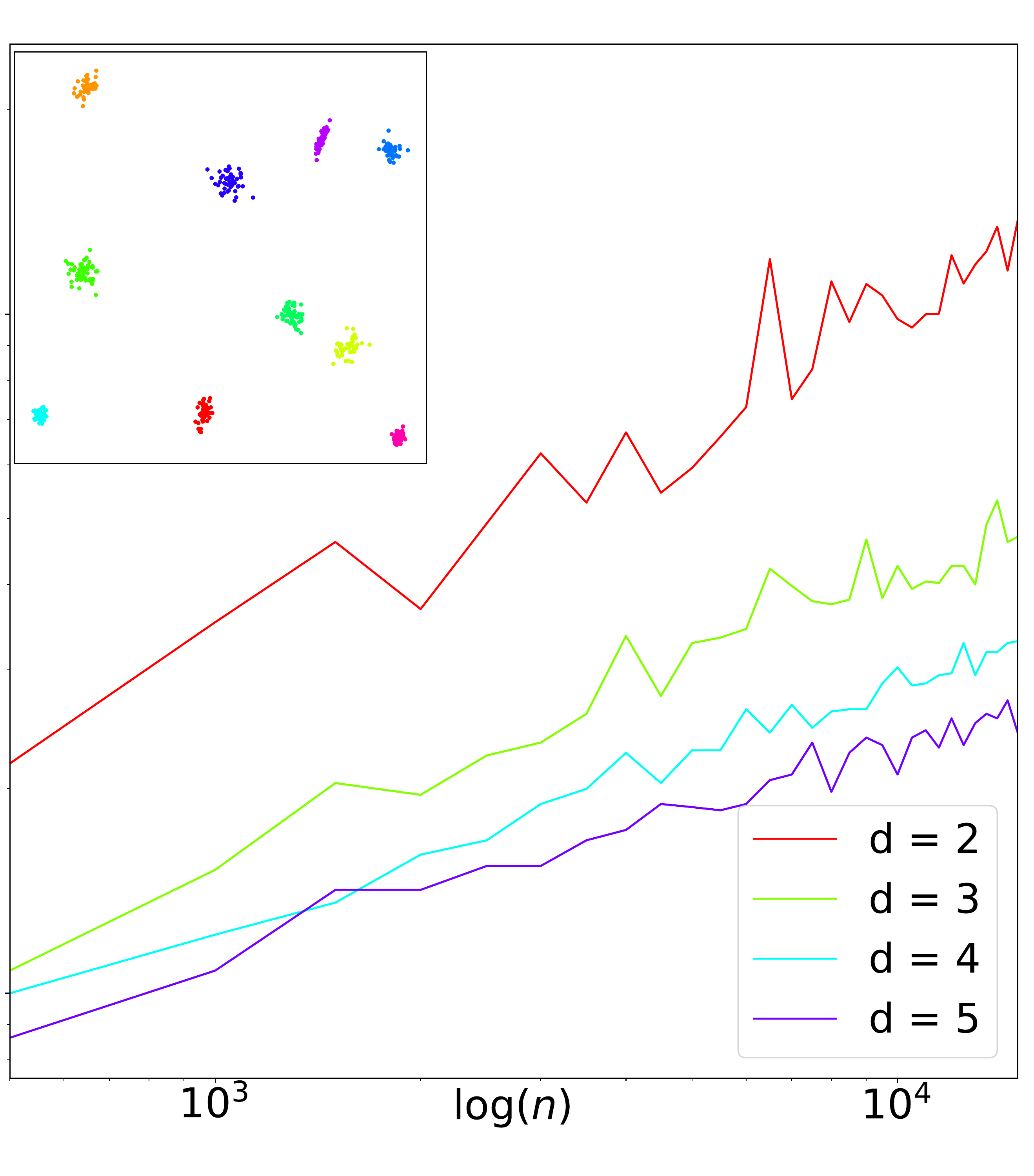}
	}
\end{subfloat}
\begin{subfloat}[]{\label{fig:clustering_forest}
	\includegraphics[height = 19em]{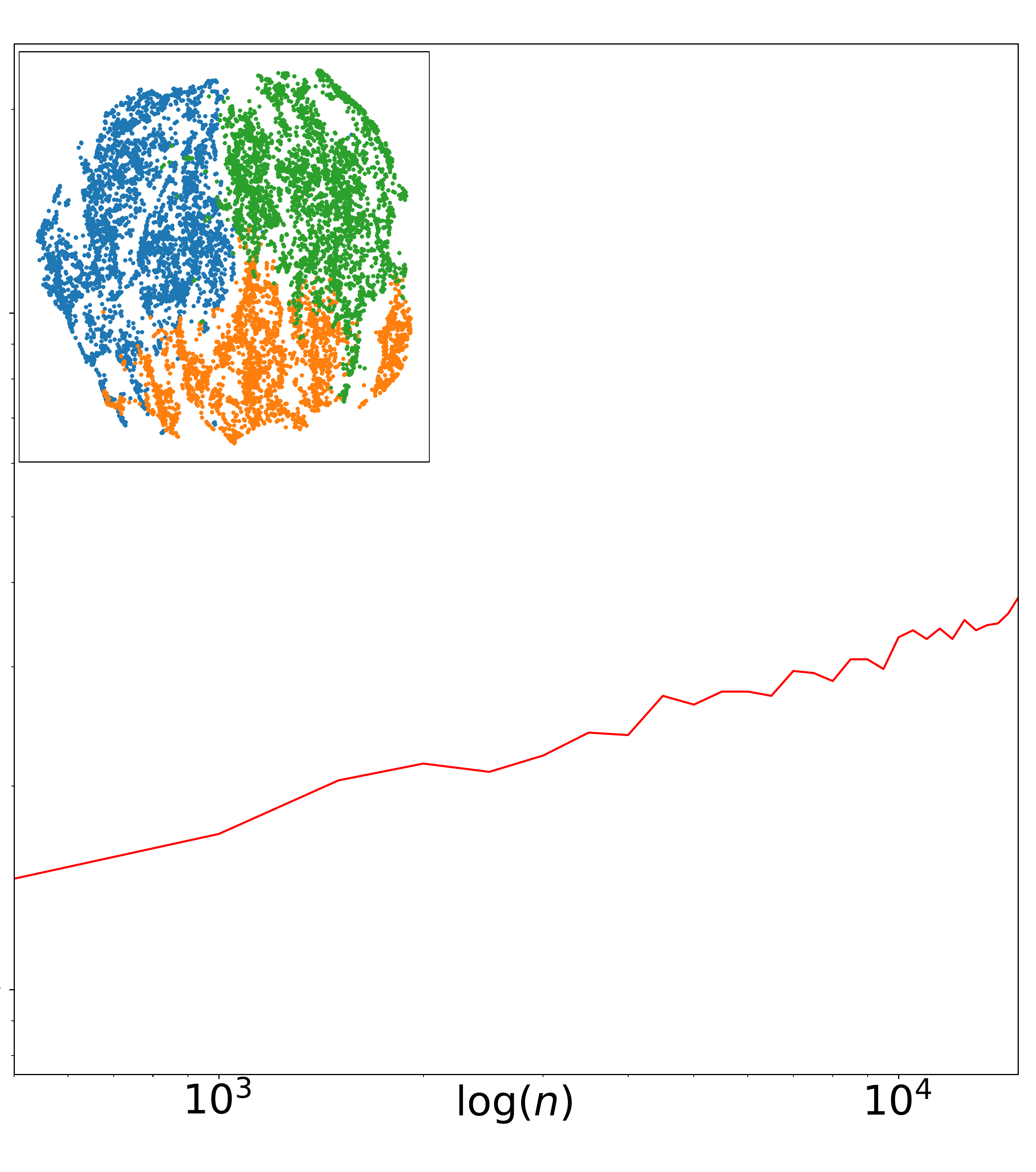}
	}
\end{subfloat}
\caption{Heights of trees. 
We plot the maximum height of any tree in the graphs of nearest-highers, $H$, for different types of datasets (we show the average over five runs). 
Both axis are shown in logarithmic scale.
For plots (a) and (b) we artificially generate uniform and Gaussian datasets (respectively) for several dimensions $d$ with varying number of elements $n$. In the upper left corners, we show examples of such datasets in two dimensions.
For plot (c), we take $n$ random samples from the Forest Cover Type dataset \cite{ForestType}.
To visualize the dataset in two dimensions, we used a t-distributed stochastic neighbour embedding \cite{tSNE}.}
\label{fig:heights}
\end{figure}

To understand how the factor $H$ scales, we can start by considering very simple data model. 
We assume that the dataset is generated by sampling $n$ points uniformly at random from a bounded region of $\R^d$.
In this case, we expect to see only one cluster that spans the entire region.
A straightforward calculation reveals that the expected nearest-neighbour distance $\expval{\mathit{nn}}$ scales as
\begin{equation}
	\expval{\mathit{nn}} = \mathcal{O} \left( \frac{1}{n^{1/d}} \right).
\end{equation}
While the nearest-higher does not always coincide with the nearest-neighbour, the nearest-higher separation is most likely not much larger than the typical nearest-neighbour distance (except for the root node).
So, we should have 
\begin{equation}
	\expval{\delta} \sim \expval{\mathit{nn}}.
	\label{eq:deltascaling}
\end{equation}
Now consider a leaf node of the graph of nearest-highers at a distance, say, $L$ from the root node.
As it probably lies close the edge of the region, $L$ characterizes the ``size'' of the cluster.
Its nearest-higher is found roughly along the direction of the centre of the cluster.
That is, the nearest-higher is at a distance $\sim \expval{\delta}$ closer to the root node.
Repeating this reasoning for all nodes in the path to the root, we conclude that
\begin{equation}
	\expval{H} 
	\sim \frac{L}{\expval{\delta}} 
	\sim \frac{L}{\expval{\mathit{nn}}}
	 = \mathcal{O} \left( n^{1/d} \right).
	 \label{eq:Hscaling}
\end{equation}
Admittedly, we have merely provided an informal argument for the expression \eqref{eq:Hscaling}, not a fully rigorous proof.
Still, we can verify this behaviour numerically.
We generate such artificial datasets by sampling uniformly from $d$-balls of radius one, for different dimensions $d$.
The results, shown in Figure \ref{fig:heights}a, confirm that indeed $\expval{H}$ scales as $n^{1/d}$.

To study how  $H$ behaves in more general settings, we consider two other types of datasets:
\begin{itemize}
\item \emph{Well-separated, Gaussian-shaped clusters (Figure \ref{fig:heights}b).} 
For different values of $d$, we pick ten centroids at random in range $[0, 100]^d$, associating to each a randomly chosen covariance matrix. 
We then generate artificial datasets by drawing from the corresponding Gaussian distributions.
With high probability, the clusters will be well-separated.
\item \emph{Real-world dataset (Figure \ref{fig:heights}c).}
We randomly sample entries from the Forest Cover Type dataset \cite{ForestType}.
This dataset originally contains fifty-five features, both numerical and categorical, conveying information about the Roosevelt National Forest in Colorado, namely, tree types, shadow coverage, distance to nearby landmarks, soil type, and local topography.
We preprocess the dataset by numerically encoding the categorical data, and then rescaling the numerical variables such that each has mean zero and variance one.
We define the clustering distance as the Euclidean distance between the first ten principal components of the data. 
\end{itemize}
For both cases, we observe that
\begin{equation}
	\expval{H} \sim n^{1 / \deff},
\end{equation}
for some parameter $\deff$.
We interpret $\deff$ as an ``effective dimension'' of the dataset, which can be smaller than the number of features of the data.
For example, for the Gaussian datasets a simple polynomial fit reveals $\deff = 1.94, 2.36, 2.80, 3.22$ for $d = 2,3,4,5$, respectively.
For the Forest Cover Type dataset, we find $\deff = 3.71$.
An interesting research question (outside the scope of the present article) would be to properly understand how this effective dimension arises from the structure of the data.

For the cases considered where the datasets had more than two features, we have verified that the effective dimension was greater than two, entering the regime where our density peak decision clustering algorithm shows a quantum speedup.
This is evidence that our quantum algorithm could be suitable for high-dimensional, real-world problems.
Our conclusions are summarized in the following claim.
\begin{claim}
Let $D$ be a dataset of $n$ elements and let the maximum height of any tree in the graph of nearest-highers scale as $\mathcal{O}\left(n^{1 / \deff} \right)$ for some parameter $\deff$.
Then, for any two elements $x_i, x_j \in D$, we can solve density peak decision clustering with constant error probability with quantum query complexity
\begin{equation}
\tilde{\mathcal{O}} \left( n^{3/2 + 1 / \deff} \right).
\end{equation}
In particular, if $\deff > 2$, quantum query complexity is better than the classical query complexity $\mathcal{O} \left( n^2 \right)$.
\end{claim}

We would like to stress that our speedup shows a dependence on a geometric property of the dataset that, to the best of our knowledge, has not yet been seen in the literature.
While it is a common idea that quantum speedups in machine learning may rely on the structure of the dataset, it is usually hard to rigorously characterize the necessary structure.
In contrast, in this work we were able prove that we have speedup if $H$ scales better than $\sqrt{n}$ (or, in other words, if the effective dimension is larger than $2$).

\section{Experimental implementation \label{sec:experiment}}

\begin{figure}[t]
\centering
\begin{subfloat}[]{\label{fig:toy_dataset}
	\includegraphics[width=0.48\linewidth]{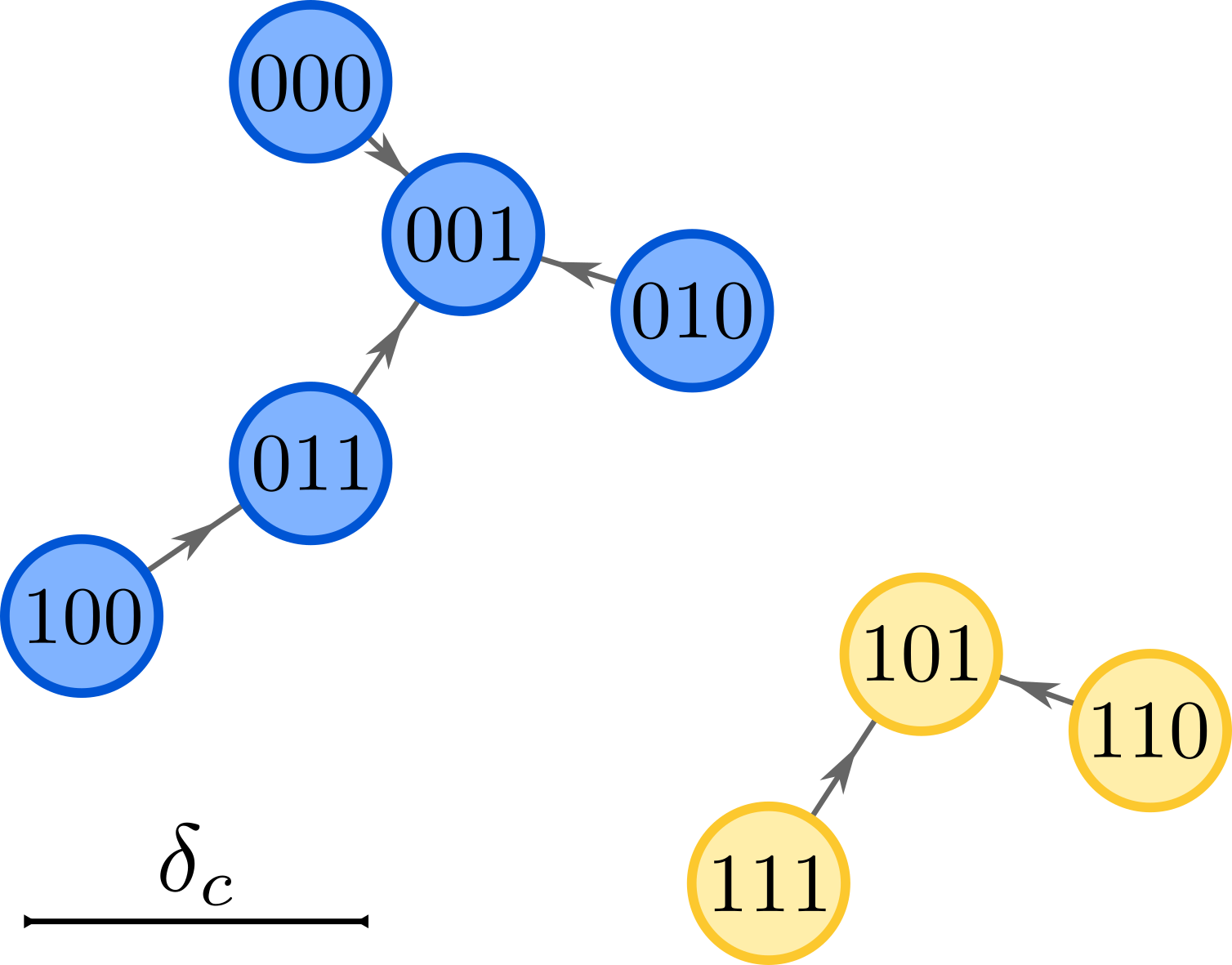}
	}
\end{subfloat}
\begin{subfloat}[]{\label{fig:simul_results}
	\includegraphics[width=0.48\linewidth]{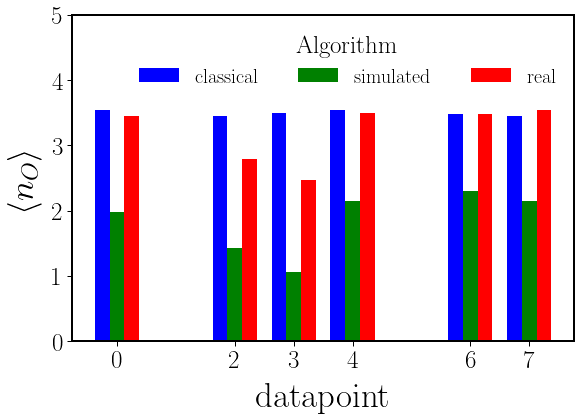}
	}
\end{subfloat}
\caption{Experimental implementation of toy problem. Figure \ref{fig:toy_dataset} shows toy dataset of 8 points. The two colours (blue and yellow) denote the two distinct clusters in the dataset. The arrows denote the nearest-higher (except for the elements $1$ and $5$ which are the cluster centers). 
In plot \ref{fig:simul_results} we plot the average number of iterations, or average number of oracle calls $\langle n_O \rangle$, to find the nearest-higher for each of the points in the dataset. We reported a classical random search strategy, an ideal simulated quantum minimum finding subroutine and the implementation of the same quantum subroutine in a real IBM quantum processor. The average is taken over $1000$ consecutive runs of each strategy. The datapoints reported in the x-axis of the bar graph are the decimal representation of the binary points in the dataset (e.g. datapoint $3$ is the element $011$ of the dataset), notice that datapoints $1$ and $5$ are missing from the bar graph as they are the cluster centers and their number of iterations is fixed.}
\end{figure}

In this section, we test the proposed quantum density peak clustering on a real quantum processor.
Specifically, we implement the main quantum routine, minimum finding. 
Since we are limited by the size of the devices, we solve a synthetic clustering problem involving just eight elements -- see the dataset depicted in Figure \ref{fig:toy_dataset}. 
We can encode each element of the dataset in $\lceil \log(8) \rceil = 3$ qubits, being suitable to run on NISQ machines. 
The first problem to address is implementing the oracle in a suitable manner.

Given the density $\rho_i$ computed for each $i \in \{0, \ldots, 7\}$ in our synthetic dataset, consider the function introduced in Eq.~\eqref{eq:ffunction}. The quantum minimum finding calls an oracle $O_{i,j}$ that implements the Boolean function
\begin{equation}
F_{i,j}(k)=
\begin{cases}
0, \text{ if } f_i(k) \geq f_i(j) \\
1, \text{ if } f_i(k) < f_i(j)
\end{cases}
\end{equation}
as
\begin{equation}
O_{i,j} \vert k \rangle=
\begin{cases}
+\vert k \rangle, \text{ if } F_{i,j}(k)=0 \\
-\vert k \rangle, \text{ if } F_{i,j}(k)=1
\end{cases}.
\label{eq:experimental_oracle}
\end{equation}
Present-day quantum machines do not have qRAM access, nor can perform complex arithmetic operations such as computing densities.
So, we classically pre-compute $F_{i,j}(k)$ for every values of $i$, $j$, and $k$ and construct a circuit for each oracle $O_{i,j}$ following scheme outlined in \cite{figgatt2017complete}.
While this is not how the algorithm is meant to be implemented in practice (cf. section \ref{sec:qalg}), we believe that this approach is suited for the purposes of a proof-of-concept.

At this point, we have all the ingredients to start the quantum nearest-higher search. 
Given a point $i$ we start by selecting a random threshold $j$ and call the oracle dictionary $F_{i,j}$. 
We then mark the states that have $F_{i,j}(k)=1$ following the same marking scheme outlined in \cite{figgatt2017complete} and apply the amplitude amplification subroutine. 
At the end, we measure a state $\vert k \rangle$ that is selected as the new threshold $j'=k$ if $f_i(k) \geq f_i(j)$, otherwise a new threshold $j'$ is selected at random. This procedure is iterated until the nearest higher is found for every point $i$ in the dataset. 

We benchmark this strategy against the classical method of nearest-higher search that is just a random sampling of $j$. 
In either case, the figure of merit is the average number of oracles calls before finding the nearest-higher, $\langle n_O \rangle$, i.e., the quantum query complexity.

In Figure \ref{fig:simul_results} we show the results for $\langle n_O \rangle$ taken over $1000$ run of the algorithm by using both the classical strategy and the quantum routine. 
As expected, the classical nearest-higher algorithm (i.e. random search) always takes on average  iterations $\sim 3.5$ (blue bars in Figure \ref{fig:simul_results}), irrespectively of the point as one would expect given the size of the dataset. 

Regarding the quantum search, we first used the Qiskit package \cite{Qiskit} to simulate the algorithm without noise. 
In general, we can have multiple rounds of amplitude amplification in each step of our quantum minimum finding subroutine. 
However, here we opted by always running just one round.
The simulations (green bars in Figure \ref{fig:simul_results}) clearly demonstrate quantum speedup, as the average number of oracle calls before convergence is lower than in the classical case for all points. We can observe that there are some points with speedups more pronounced than others as, for each specific datapoint, this depends on the number of states marked by the oracle.

Finally, we ran our Qiskit program on a real quantum computer by IBM, the ibm-perth 7-qubit processor (see red bars in Figure \ref{fig:simul_results}). 
In this case, the analysis is more subtle because of the real-world noise on top of our ideal quantum algorithm. 
Indeed, for points whose oracles mark several states, the depth search circuits is quite large, making the computations more sensible to noise.
On the other hand, there are some points for which the number of states marked by the oracle is low and thus the circuit is small enough that we can still observe an advantage over the classical case (even if it is less pronounced than in the ideal quantum simulation due to noise).
This observation is very relevant as it is generally difficult to observe such quantum advantages running quantum machine learning subroutines on real hardware, even if for a toy problem as the one explored here. 
We can thus hope that, with some improvement in coherences and maybe error correcting codes being built in real quantum processors, we can start observing some advantages for real-world problems.
\section{Conclusions \label{sec:conclusion}}

In this work, we have introduced a quantum version of the density peak clustering algorithm, specifically aimed at its decision version. 
Our proposed algorithm builds upon the well-known quantum minimum finding algorithm,  giving us the possibility of computing the query complexity of quantum density peak clustering. 
Indeed, while the classical query complexity of density peak clustering is $\mathcal{O} \left( n^2 \right)$, we prove that our proposed quantum algorithm has complexity $\mathcal{O}\left( n^{3/2 + 1 / \deff} \right)$, for a parameter $\deff$ that depends on the structure of the dataset. 
For values of $\deff>2$, we have a quantum speedup, albeit a modest one. 
This provable dependence of the complexity on this geometric property of the dataset constitutes by itself an notable result. 
Indeed, while it is widely accepted that quantum speedups for machine learning may depend on the structure of the data, it is often difficult to precisely characterize this dependence.
As discussed in section \ref{sec:height}, in our case we interpret the parameter $\deff$ as an ``effective dimension'' of the dataset, making quantum density peak clustering specially suited for high-dimensional problems. 
The successful implementation of a toy problem in a real quantum computer and the observation of an advantage, even in the presence of the noise typical of a NISQ device, hints at the concrete possibility of exploiting the capability of quantum density peak clustering in near-term quantum computers.

To conclude, we would like to raise two points that are relevant not only for the quantum density peak clustering, but also for the quantum machine learning community at large. 
The first one regards the efficient implementation of the classical computation routines, which in this work were included in the oracles.
For example, while comparing two distances consumes $\mathcal{O}(1)$ time, the overhead of this calculation prohibits its implementation on present-day quanutm hardware. 
The second point is to understand how the effective dimension of the dataset $\deff$ scales in general for an arbitrary dataset (and if we can always define an effective dimension given the scaling of nearest neighbors). 
Going even deeper, one could ask if $\deff$ represents some fundamental property of the data and its structure (and can thus be exploited further) or if it is just a scaling parameter.
We believe that answering these questions could be crucial for quantum density peak clustering and its future as a viable clustering algorithm for quantum machine learning on near-term quantum devices.

\section*{Acknowledgments}

We would like to thank Bruno Coutinho for his valuable insights on the graph of nearest-highers.
We would also like to thank Diogo Cruz, Akshat Kumar, Jo\~{a}o Moutinho, Sagar Pratapsi, and Mathieu Roget for fruitful discussions.

Furthermore, we acknowledge the support from FCT, namely through project UIDB/50008/2020 and UIDB/04540/
2020.
DM acknowledges the support from FCT through scholarship 2020.04677.BD.

\end{document}